\documentclass[letter]{aa}
\usepackage{graphics,epsfig,txfonts,xspace}
\usepackage{natbib}
\bibpunct{(}{)}{;}{a}{}{,}

\newcommand{\ergcm}[1]{$\times 10^{#1}$ erg cm$^{-2}$ s$^{-1}$}

\newcommand{\ergs}[1]{$\times 10^{#1}$ erg s$^{-1}$}
\newcommand{\oergs}[1]{$10^{#1}$ erg s$^{-1}$}
\newcommand{\hcm}[1]{$\times 10^{#1}$ cm$^{-2}$}

\newcommand{\expo}[1]{$\times 10^{#1}$}
\newcommand{\oexpo}[1]{$10^{#1}$}
\newcommand{\kms}{km s$^{-1}$}

\newcommand{\Hone}{\ion{H}{i}}

\newcommand{\SII}{[\ion{S}{ii}]}
\newcommand{\OIII}{[\ion{O}{iii}]}
\newcommand{\Halp}{H${\alpha}$}
\newcommand{\ltsima}{$\buildrel < \over \sim$}
\newcommand{\lsim}{\lower.5ex\hbox{\ltsima}}
\newcommand{\gtsima}{$\buildrel > \over \sim$}
\newcommand{\gsim}{\lower.5ex\hbox{\gtsima}}

\newcommand{\fdot}{$\dot{\rm f}$}
\newcommand{\Pdot}{$\dot{\rm P}$}
\newcommand{\rahour}{\hbox{\ensuremath{^{\rm h}}}}
\newcommand{\ramin}{\hbox{\ensuremath{^{\rm m}}}}

\newcommand{\sxp}{\mbox{SXP\,1062}\xspace}
\newcommand{\cxoj}{CXO\,J012745.97$-$733256.5\xspace}

\begin{document}
 
\title{SXP\,1062, a young Be X-ray binary pulsar with long spin period
       \thanks{Based on observations with 
               XMM-Newton, an ESA Science Mission with instruments and contributions 
               directly funded by ESA Member states and the USA (NASA)}}
\subtitle{Implications for the neutron star birth spin}

\author{F.~Haberl\inst{1} \and R.~Sturm\inst{1} \and M.\,D.~Filipovi{\'c}\inst{2} \and W.~Pietsch\inst{1} \and E.\,J.~Crawford\inst{2}}

\titlerunning{SXP\,1062, a young Be X-ray binary pulsar with long spin period}
\authorrunning{Haberl et al.}
 
\institute{Max-Planck-Institut f\"ur extraterrestrische Physik,
           Giessenbachstra{\ss}e, 85748 Garching, Germany, \email{fwh@mpe.mpg.de}
	   \and
           University of Western Sydney, Locked Bag 1797, Penrith South DC, NSW1797, Australia
	   }
 
\date{Received 1 November 2011 / Accepted 1 December 2011}
 
\abstract
{The Small Magellanic Cloud (SMC) is ideally suited to investigating the recent star formation history from X-ray source population studies. 
It harbours a large number of Be/X-ray binaries (Be stars with an accreting neutron star as companion), and the supernova remnants 
can be easily resolved with imaging X-ray instruments.}
{We search for new supernova remnants in the SMC and in particular for composite remnants with a central X-ray source.}
{We study the morphology of newly found candidate supernova remnants using radio, optical and X-ray images and investigate their X-ray spectra.}
{Here we report on the discovery of the new supernova remnant around the recently discovered Be/X-ray binary pulsar CXO\,J012745.97$-$733256.5 = \sxp in 
radio and X-ray images. 
The Be/X-ray binary system is found near the centre of the supernova remnant, which is located at the outer edge of the eastern wing of the SMC. 
The remnant is oxygen-rich, indicating that it developed from a type Ib event.
From XMM-Newton observations we find that the neutron star with a spin period of 1062 s (the second longest known in the SMC) shows a 
very high average spin-down rate of 0.26 s per day over the observing period of 18 days.}
{From the currently accepted models, our estimated age of around 10000$-$25000 years for the supernova remnant is not long enough to spin down 
the neutron star from a few 10 ms to its current value. Assuming an upper limit of 25000 years for the age of the neutron star
and the extreme case that the neutron star was spun down by the accretion torque that we have measured during the XMM-Newton observations since
its birth, a lower limit of 0.5~s for the birth spin period is inferred.  
For more realistic, smaller long-term average accretion torques our results suggest that the neutron star was born with a correspondingly
longer spin period. This implies that neutron stars in Be/X-ray binaries with long spin periods can be much younger than currently anticipated.}


\maketitle

 \section{Introduction}

One of the most cataclysmic events in the universe is the explosion of a massive star as a supernova, which can create a neutron star (NS).
NSs are thought to be born rapidly spinning with rotation periods of a few 10 ms. Their rotation is first slowed down by magnetic dipole 
braking and then by the propeller effect when the NS is in a binary star system and mass loss from the companion star begins
\citep[for a review see e.g.][]{1991PhR...203....1B}. 
When rotating slowly enough, accretion onto the NS sets in \citep{1978ApJ...223L..83G} and the system can be detected as an X-ray binary. 
The spin evolution of the NS in a high-mass X-ray binary (HMXB) system depends on the initial magnetic field strength of the NS and 
on the mass accretion rate \citep{1998MNRAS.299...73U}. 

Be/X-ray binary (BeXRB) systems are a subgroup of HMXBs with an NS accreting matter from the circumstellar disk 
of a Be star \citep{2011Ap&SS.332....1R}. 
The Small Magellanic Cloud (SMC) harbours an extraordinarily large number of BeXRBs
\citep{2005MNRAS.356..502C,2005MNRAS.362..879S,2008A&A...489..327H}. 
These are believed to have been created during a burst of star formation about 42 million years ago
\citep{2010ApJ...716L.140A}.
For more than fifty BeXRBs in the SMC the NS spin period is known from the detection of coherent pulsations in their X-ray flux. 
The spin periods range from 2.16~s \citep[XTE\,J0119-731 = SXP\,2.16; ][]{2003IAUC.8064....4C} 
to 1320 s \citep[RX\,J0103.6-7201 = SXP\,1323; ][]{2005A&A...438..211H} with a bimodal structure 
indicated in their distribution \citep{2011arXiv1111.2051K}.

From ROSAT HRI data, \citet{1994AJ....107.1363H} proposed two BeXRBs within supernova remnants (SNRs) in the SMC. For one of them, located 
in the direction of the SNR IKT\,21 \citep{2004A&A...421.1031V}, pulsations of 345~s were discovered by 
\citet[][SAX J0103.2-7209]{2000ApJ...531L.131I}.
The other - still unconfirmed - candidate BeXRB in the direction of IKT\,25 was not detected by XMM-Newton or Chandra.
In both cases it is not clear whether the BeXRB (or candidate) is physically associated with the SNR because they are not located near the centre of the 
remnant and the high number of BeXRBs in the bar of the SMC make chance coincidences likely. 
The SNR, which was created by the supernova explosion, is expected to have faded beyond detectability before 
accretion onto the NS starts. 
Finding a BeXRB associated to a SNR would therefore suggest a much faster spin-down than predicted by the standard models, or else the 
NS star was born with slow rotation. Here, we present the discovery of an SNR in radio and X-rays around the BeXRB pulsar \cxoj\ = \sxp. 
The discovery of the new SMC pulsar and its possible association with a shell nebula around it has recently been reported by
\defcitealias{Henault}{HB11}\citet[][ hereafter \citetalias{Henault}]{Henault}.

\begin{figure*}
\begin{center}
\resizebox{0.92\hsize}{!}{\includegraphics[clip=]{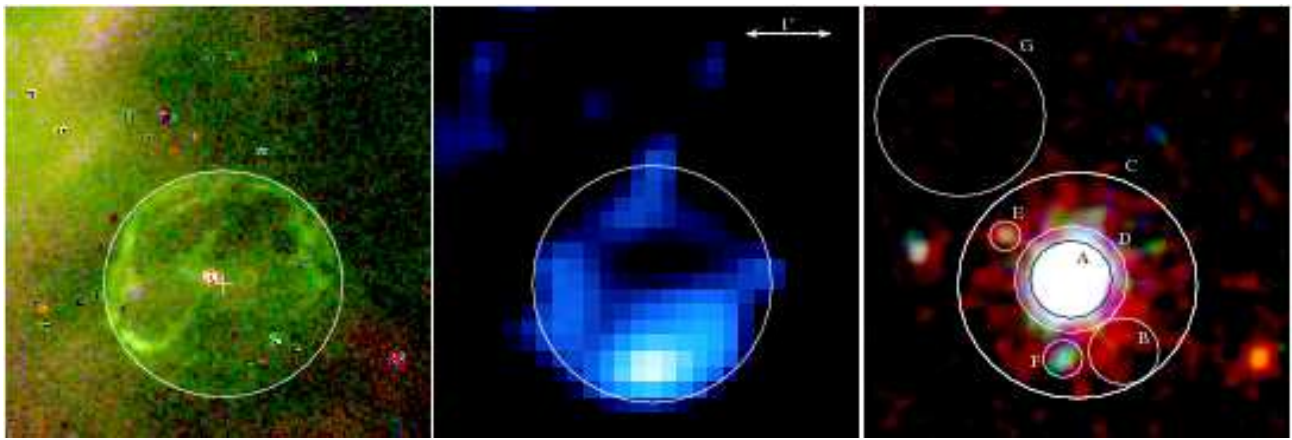}}
\caption{Images of the region around the BeXRB SXP\,1062.
{\it Left:}
            Continuum-subtracted MCELS images. Red, green, and blue correspond to H$\alpha$, \OIII\ and \SII. 
            The cross marks the centre of the white circle (R.A. = 01\rahour27\ramin44\fs15, Dec. = $-$73\degr33\arcmin01\farcs6, J2000), 
            which indicates the estimated size of the SNR in all images. 
            The Be star counterpart of \sxp is visible close to the cross.
{\it Middle:}
            MOST 36\,cm radio image, logarithmically scaled from 0.5 to 3 mJy. 
{\it Right:}
           Combined, instrument background-subtracted EPIC-pn and MOS colour image obtained from the XMM-Newton observations. Red, green, and blue denote X-ray
           intensities in the 0.2$-$1.0, 1.0$-$2.0, and 2.0$-$4.5 keV bands. 
           The images from the individual bands were adaptively smoothed with a Gaussian $\sigma$ of 4\arcsec\ (at high intensities) to 20\arcsec\ (low intensities).
           Circles indicate the extraction regions for the X-ray spectra:
           Source region for SXP\,1062 (A, 29\arcsec\ radius) and associated background (B, 25\arcsec).
           For the SNR spectrum (C, 84.5\arcsec), regions D (40\arcsec), E (11\arcsec), and F (14\arcsec) were excluded and region G (60 \arcsec) was used to estimate the background. 
           Regions E and F were excluded because they contain possible point sources in the direction of the SNR as found in source detection runs on the EPIC images.
       }
\label{fig-ima}
\end{center}
\end{figure*}

\section{Multi-wavelength observations and data analysis}

\subsection{Optical}

The Magellanic Cloud Emission Line Survey\footnote{MCELS: http://www.ctio.noao.edu/$\sim$mcels/} was carried out from the 
0.6~m University of Michigan/CTIO Curtis Schmidt telescope, equipped with an SITE 2048$\times$2048 CCD. This resulted in a field of 
1.35\degr\ $\times$ 1.35\degr\ with a pixel scale of 2.4\arcsec\ $\times$ 2.4\arcsec. The Magellanic Clouds were 
mapped in narrow wavelength bands corresponding to \Halp, \OIII, and \SII, together with matching red and green continuum bands.
All the images were continuum-subtracted, flux-calibrated, and assembled into mosaic images. 
A composite image in the three bands of the region around \sxp is presented in Fig.~\ref{fig-ima}, revealing a shell-type nebula 
dominated by the strong \OIII\ line as also discussed by \citetalias{Henault}.

\subsection{Radio-continuum}

The field of \sxp was observed with the Australia Telescope Compact Array (ATCA) on 2009-01-05 with array configuration 6C, and on 2009-02-05 in 
array configuration EW352 (ATCA project C1869). The field was observed at wavelengths of 20\,cm and 13\,cm.
Each session was carried out 
in ``snap-shot'' mode, totalling $\sim$3.5 hr of integration time over a 12 hr period. These data were combined with the relevant parts of the data from 
\citet{2011SerAJ.182...43W} and \citet{1995A&AS..111..311F} to improve $u$-$v$ plane coverage. 
The 13\,cm data proved unusable due to radio frequency interference. The 20\,cm data was useful only when long baselines were excluded thus the final 
image has a resolution of $145\arcsec\times 131\arcsec$ with  an r.m.s. noise of 0.5~mJy/beam. We report an integrated flux density of 6\,mJy at 20\,cm.

Using the image of \citet{1993LNP...416..167Y}, we measured an integrated flux density of 9\,mJy at 36\,cm 
(see  Fig.~\ref{fig-ima}). 
Combining this measurement with the 20\,cm measurement  we derive a spectral index ($S_\nu\propto \nu^\alpha$) of $\alpha=-0.8\pm0.4$ 
which is steeper than most SNRs in the Magellanic Clouds \citep{1998A&AS..130..421F}. We estimate that the surface brightness of this SNR is 
$1.6\times 10^{-22}$ W\,m$^{-2}$\,Hz$^{-1}$\,Sr$^{-1}$ with a luminosity of $3.4\times 10^{15}$ W\,Hz$^{-1}$, assuming a distance of 60~kpc and diameter of $166\arcsec$. 
This makes it the faintest known radio SNR in the SMC \citep{2005MNRAS.364..217F,2008A&A...485...63F, 2011A&A...530A.132O}.

\subsection{X-rays}

Chandra (between 2010-03-31 and 2010-04-29) and XMM-Newton (2010-03-25 to 2010-04-12) observed the field around the star-forming region 
NGC\,602 and revealed the new BeXRB \sxp in the eastern wing of the SMC \citepalias{Henault} with a spin period of 1062 s, the second longest known 
in the SMC after SXP\,1323. Here, we present the analysis of the four XMM-Newton observations 
(observation IDs 0602520401, 0602520201, 0602520301, and 0602520501) with durations of 66.3, 119.0, 91.2, and 55.0 ks and spread over 18 days.

\subsubsection{Morphology}

Following the analysis of the SMC survey with XMM-Newton (Haberl et al. in preparation, Sturm et al. in preparation), 
we produced X-ray images from the 
combined EPIC-pn \citep{2001A&A...365L..18S} and EPIC-MOS \citep{2001A&A...365L..27T} data in the energy bands 
0.2$-$1.0 keV, 1.0$-$2.0 keV, and 2.0$-$4.5 keV. Our analysis of the EPIC data used the tools from the 
XMM-Newton SAS package version 11.0.1\footnote{Science Analysis Software (SAS), http://xmm.esac.esa.int/sas/}. 
To optimise the sensitivity for faint diffuse X-ray emission we removed intervals of 
high background caused by soft protons and subtracted the contribution of the detector particle background from the images. 
For EPIC-pn also the out-of-time events recorded during CCD readout were subtracted. 
The final colour image of the region around \sxp is presented in Fig.~\ref{fig-ima}.
Although the emission from \sxp dominates the hard energy bands, soft extended emission is clearly seen around the X-ray pulsar. The extent 
of this emission is similar to the radio and MCELS images. We also checked the Chandra images for the soft emission, 
but the sensitivity at low energies is insufficient for the Chandra ACIS-I instrument to allow a significant detection.

\subsubsection{Spectral analysis}

We extracted EPIC spectra from the X-ray pulsar and the soft emission region, merging the data from all four observations. The extraction 
regions are indicated in Fig.~\ref{fig-ima} and described below. No MOS1 data is available because the source was located on CCD 6, which 
is no longer operative. 
For the spectra we selected single and double-pixel events (PATTERN 0-4) for pn and single to quadruple events (PATTERN 0-12) for 
MOS with FLAG=0 to avoid bad pixels. Applying the same background flare screening as for the images resulted in net exposures for 
the spectra of 206.8 ks for pn and 236.7 ks for MOS2. 

\begin{figure}
\begin{center}
\resizebox{0.85\hsize}{!}{\includegraphics[clip=,angle=-90]{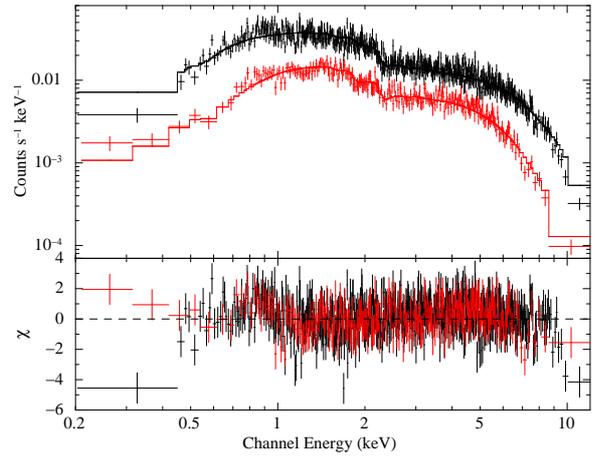}}
\caption{EPIC-pn (black) and MOS2 (red) spectra of \sxp combining the data from the four XMM-Newton observations. The best-fit 
ower-law model is plotted as histogram, and the lower panel shows the fit residuals.}
\label{fig-sxpspec}
\end{center}
\end{figure}

To extract the spectra of \sxp we used a circular region and a very close background region inside the soft emission region to remove the 
contribution of this component.
We fitted the pn and MOS2 simultaneously with a power-law, attenuated by two absorption components. The first accounts for the Galactic 
foreground absorption with a fixed column density of 6\hcm{20} and solar abundances \citep{2000ApJ...542..914W}, while the second models 
the absorption along the line of sight within the SMC and local to the source with free column density and reduced abundances 
of 0.2 for elements heavier than helium \citep{1992ApJ...384..508R}. This resulted in an acceptable fit with reduced $\chi^2$ of 
1.12 for 377 degrees of freedom (see Fig.~\ref{fig-sxpspec}). The best-fit value for the column density in the SMC 
was $1.8\pm0.2$ \hcm{21} and the power-law photon index $0.74\pm0.02$, values typical of BeXRBs in the SMC \citep{2008A&A...489..327H}. 
The observed average flux (0.2$-$10 keV) during the XMM-Newton observations was 1.4\ergcm{-12} (as obtained from the pn spectrum),
which corresponds to a source luminosity corrected for absorption of 6.3\ergs{35} (assuming a distance of 60 kpc). 
The flux increased after the first observation and finally dropped by 27\% in the last observation 
(1.33\ergcm{-12}, 1.53\ergcm{-12}, 1.54\ergcm{-12} and 1.13\ergcm{-12} in chronological order), which could indicate that we 
witnessed a typical type I outburst \citep{2001A&A...377..161O}.

\begin{figure}
\begin{center}
\resizebox{0.85\hsize}{!}{\includegraphics[clip=,angle=-90]{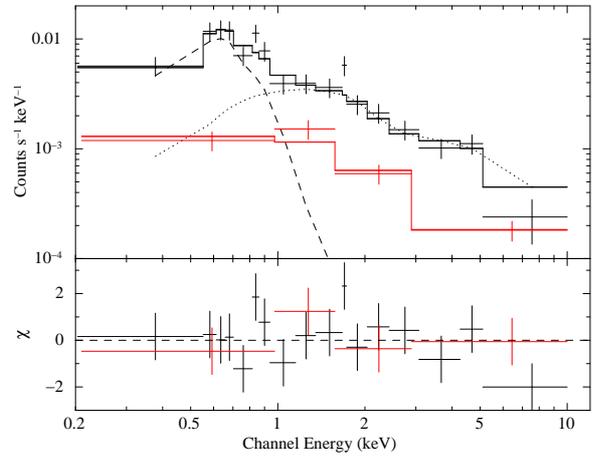}}
\caption{EPIC spectra as in Fig.~\ref{fig-sxpspec} extracted from the soft emission region around \sxp. 
The best-fit model is composed of a thermal (dashed line) and a power-law (dotted) component (see text).}
\label{fig-snrspec}
\end{center}
\end{figure}

To investigate the soft extended emission, we extracted the EPIC spectra from a ring around the position of \sxp. The relatively 
wide wings in the point spread function of the XMM-Newton telescopes means a contribution from the X-ray pulsar to the soft emission cannot 
be avoided. Therefore, we modelled the spectrum of the diffuse emission with a power-law component in addition to a thermal component. We kept
the SMC absorption and photon index for the power-law fixed at the values derived from the spectra of \sxp, allowing only a free 
normalisation. For the thermal component we used the plasma emission model {\tt mekal} (also with SMC abundances as for the absorption) 
available in \textsc{XSPEC} \citep{1985A&AS...62..197M} with SMC absorption (free in the fit and different to that of the power-law component). 
The resulting best-fit parameters (reduced $\chi^2$ = 1.22 for 16 degrees of freedom) are 7.3$^{+8.8}_{-6.8}$\hcm{20} for the SMC absorption
and 0.23$\pm$0.05 keV for the temperature. The observed flux (0.2$-2$ keV) of 1.1\ergcm{-14} makes it the X-ray faintest SNR known in 
the SMC \citep[for other faint remnants see ][]{2008A&A...485...63F}. The spectra with the best-fit model are presented in Fig.~\ref{fig-snrspec}. 
Unfortunately, the low-quality statistics of the spectra do not allow a more detailed comparison of different models for the thermal 
spectral component. 

The \Hone\ column density of the SMC in the direction of \sxp\ is measured to 2.1\hcm{21} \citep{1999MNRAS.302..417S}. The absorption 
inferred from the X-ray spectrum of \sxp is similar, while for the soft emission region we can only place an upper limit of 
$\sim$1.6\hcm{21}. The fact that BeXRBs show source intrinsic absorption and the large uncertainty in the column density of the soft 
X-ray emission spectrum do not allow deriving any constraints on the radial position of the objects within the SMC. 
Nevertheless, the two column densities inferred from the X-ray spectra of the BeXRB and the diffuse X-ray emission region (the SNR) are 
consistent with a location of the BeXRB within the SNR.

\subsubsection{Timing analysis of \sxp}

\begin{figure}
\begin{center}
\resizebox{0.99\hsize}{!}{\includegraphics[clip=,angle=-90]{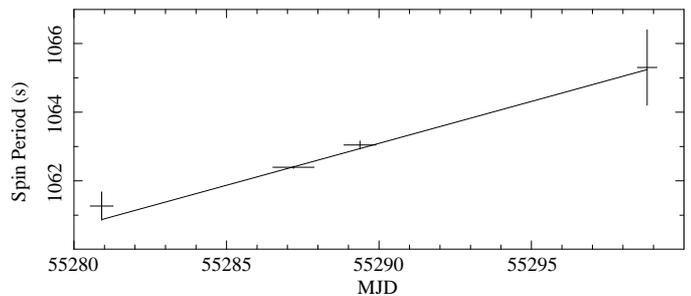}}
\caption{Evolution of the spin period of \sxp as obtained from the combined EPIC-pn and EPIC-MOS2 data of the four XMM-Newton observations. 
Spin periods and 1-$\sigma$ errors were derived with the Bayesian odds ratio method.}
\label{fig-spin}
\end{center}
\end{figure}

We performed a timing analysis of the EPIC light curves using the Bayesian approach \citep{1996ApJ...473.1059G} and a 
Rayleigh Z$^2$ test for one harmonic \citep{2002A&A...391..571H,1983A&A...128..245B} around the known 
periodicity of 1062~s. To investigate a possible evolution of the spin period we first analysed each XMM-Newton observation separately. 
The obtained spin periods are plotted versus time in Fig.~\ref{fig-spin} and show a steady spin-down trend over the observing 
interval of about 18 days. A linear fit to the spin period evolution (line in Fig.~\ref{fig-spin}) results in a period change \Pdot\ of 
2.8$^{+0.8}_{-0.7}$\expo{-6} s s$^{-1}$ (0.24$^{+0.07}_{-0.06}$ s day$^{-1}$). 
This value was used as starting point for a phase-coherent analysis combining the EPIC-pn data of all four observations
searching a grid of different values for the spin period P and a constant \Pdot. We used the Z$^2$ test with one 
harmonic to account for the first harmonic seen in the FFT power density spectrum. This resulted in a maximum Z$_1^2$ value 
of 344.1 for P = 1061.24~s (at an epoch of MJD 55280.53, the start of the first EPIC-pn exposure) and 
\Pdot = 3.0$\pm$0.5\expo{-6} s s$^{-1}$.

\section{Discussion and conclusions}

Our multi-wavelength morphological studies of the field around the new BeXRB pulsar \sxp confirm the previously unknown SNR
proposed by \citetalias{Henault}.
Its SNR nature is further supported by its X-ray spectrum obtained from both the XMM-Newton EPIC data and its steep radio spectrum. 
The strong \OIII\ line emission suggests the SNR as oxygen-rich type \citep{2005MNRAS.360...76A}, which 
usually comes from a type~Ib event, the explosion of a massive O, B, or WR star. 

\sxp is detected close to the centre of the SNR. At the outer edge of the eastern wing of the SMC, only very few BeXRBs 
are detected, and all known SNRs are located in the bar of the SMC \citep{2007MNRAS.376.1793P,2008A&A...485...63F}.
This makes a chance coincidence of \sxp with the SNR very unlikely. 
However, the ages of SNRs and BeXRBs are expected to be rather different. While SNRs can usually be seen
for at most a few \oexpo{5} years in X-rays and radio \citep[for the methods used to estimate SNR ages see e.g.][]{2005ChJAA...5..165X},  
BeXRBs are expected to become X-ray active much later (see below).
Using the relation t$_{\rm y}$\,=\,3.8\expo{2}\,R$_{\rm pc}$(kT)$_{\rm keV}^{-1/2}$ from \citet{2005ChJAA...5..165X} 
of temperature derived from X-ray spectral modelling (0.23 keV)
and the size of the SNR ($\sim$2.5\arcmin\ in diameter, corresponding to $\sim$40 pc at a distance of 60\,kpc), we estimate 
the age of the SNR to $\sim$16 ky. If one compares the ages estimated for SNRs in the SMC by \citet{2004A&A...421.1031V} 
with their sizes as listed in \citet{2010MNRAS.407.1301B}, one finds ages of at most 25 ky for remnants with diameters
around 40-50 pc. 

We estimate the distance between the BeXRB and the centre of the SNR to about 9\arcsec, which corresponds to $\sim$8\expo{13} km 
at SMC distance. From evolutionary investigations, kick velocities around 100$-$200 \kms\ were invoked for 
neutron stars in HMXB systems \citep{1995A&A...296..691P}. Using this as maximum space velocities for the binary systems and 
neglecting a velocity component in radial direction, it would take 13$-$26 ky for \sxp 
to move from the SNR centre to its current position, compatible with the SNR age estimates from above, which are also
consistent with those of \citetalias{Henault}.

Statistical measurements of BeXRBs in the SMC result in lower space velocities: \citet{2005MNRAS.358.1379C} 
inferred a lower limit of 30 \kms\ from measuring the separation of BeXRBs to nearby young stellar clusters. Similarly 
\citet{2010ApJ...716L.140A} argue from the correlation between the number of BeXRBs and the local star formation rate 42 My ago at 
their current position for a maximum 
velocity of 15$-$20 \kms. Both cases should actually be regarded as minimum velocities as the BeXRB might not have been born 
in the nearest star cluster. The work of \citet{2011arXiv1111.2051K} might also suggest that BeXRB pulsars with long spin 
period received stronger kicks. In either case, using lower space velocities (and also components in radial direction) would 
correspondingly increase the SNR age estimate.

According to the model calculations of \citet{1998MNRAS.299...73U}, it seems very unlikely that an NS can be spun down 
to a period around 1000 s within a few \oexpo{4} years if it was born with a spin period around 10~ms. Of course, such model 
calculations assume simplified approximations of the conditions in a BeXRB (in particular concerning the highly variable 
accretion usually observed in such systems), but only for models with extreme conditions (high magnetic field strength for 
the NS and low accretion rates), spin periods in excess of a few 100 s are reached after more than 10$^6$ years.
If the NS in \sxp is indeed the compact remnant from the supernova explosion, which also created the SNR around it, 
then either the NS was born with a much longer spin period than 10 ms, or some other effect plays an important 
role to efficiently decelerate the rotation of the NS.

\sxp was discovered as a transient BeXRB pulsar during XMM-Newton and Chandra observations in 2010 when its
X-ray luminosity was at the level of 6\ergs{35}. The BeXRB was not active during the ROSAT PSPC observations of the 
SMC \citep{2000A&AS..142...41H}. 
Based on the longest PSPC observation from 1991-10-07 and assuming the spectral shape measured by EPIC, 
we derive an upper limit of 1.5\ergcm{-13} (0.2$-$10 keV), assuming an upper limit of 10 cts and a 
vignetting-corrected exposure of 9.7 ks at the position of \sxp. 
This is about a factor of 10 lower than the flux during the XMM-Newton observations.
While BeXRB transients generally exhibit spin-up during intervals of increased accretion
\citep[e.g.][]{1997ApJS..113..367B}, \sxp is peculiar as it showed a strong spin-down consistent with a constant rate of 
0.26 s day$^{-1}$ over 18 days. \citet{2010MNRAS.401..252C} investigated RXTE monitoring data covering $\sim$10 years to study
the spin period changes of BeXRB pulsars in the SMC over their active periods, which typically last 50$-$500 days. 
They find four 
out of 15 systems with measured short-term spin period changes, which showed spin-down during outburst. 
In Table~\ref{tab-fdot}, we summarise the pulse period changes for some X-ray binary pulsars, which show times of spin-down.
We list the frequency derivative \fdot\ = -\Pdot/P$^2$, which is directly related to the accretion torque exerted 
on the NS. We include the three SMC BeXRB pulsars with well measured values of \Pdot\ and for comparison 
Cen~X-3 (a super-giant HMXB with Roche-lobe overflow) and GX~1+4, a low-mass system with M5 III donor star. \sxp
exhibits spin-down about five times larger than the other listed SMC BeXRBs (which have remarkably similar values). It 
should be noted that the values derived by \citet{2010MNRAS.401..252C} are average values over longer periods 
of time and that intermittent intervals of spin-up \citep[as usually seen from accreting pulsars,][]{1997ApJS..113..367B} 
reduce the longer-term average.

\sxp shows a remarkably large accretion torque, similar in magnitude to that of Cen~X-3. In this HMXB intervals 
of steady spin-up and spin-down alternate, which last typically 10$-$100 days \citep{1997ApJS..113..367B}, but 
at much higher X-ray luminosities of $\sim$\oergs{38} \citep[e.g.][]{1992ApJ...396..147N}. 
Also GX~1+4 was still observed at ~\oergs{36} during its extended lowstate 
\citep{1988Natur.333..746M}, about a factor of 2 brighter than \sxp.
This demonstrates that the generally accepted model for accretion, where the torque is directly proportional to the mass accretion 
rate (hence luminosity) is probably too simple, as similar torques can be exerted, although the luminosity is 
more than a factor of 100 different. The work of \citet{1997ApJS..113..367B} suggests that disk-accreting pulsars 
are subject to instantaneous torques with similar magnitude but opposite sign. Different average long-term 
spin-up or spin-down values of individual pulsars would then be the result of different time scales for 
reversals between spin-up and -down.

\sxp most likely also shows intervals of alternating spin-up and spin-down, which would reduce the magnitude of the 
long-term spin change. Assuming the extreme case that the NS was spun down with -2.6\expo{-12} Hz s$^{-1}$ 
(-8.4\expo{-5} Hz y$^{-1}$) over its whole life of 25 ky (the maximum age) an upper limit of 2 Hz for 
the total spin-down is derived, which corresponds to a lower limit of 0.5~s for the spin period at birth. 
For a more realistic, smaller long-term average accretion torque a correspondingly longer birth spin period is expected. 
Therefore, if the NS in \sxp is indeed the compact remnant from the supernova explosion that created the SNR, our 
results show that the NS was most likely born with a spin period much longer than a few tens of ms as generally 
adopted for an NS at birth.

\begin{table}
\caption[]{X-ray binary pulsars with periods of spin-down.}
\begin{center}
\begin{tabular}{lccl}
\hline\hline\noalign{\smallskip}
\multicolumn{1}{c}{Object} &
\multicolumn{1}{c}{Pulse} &
\multicolumn{1}{c}{\fdot} &
\multicolumn{1}{c}{Reference} \\
\multicolumn{1}{c}{name} &
\multicolumn{1}{c}{period [s]} &
\multicolumn{1}{c}{[Hz s$^{-1}$]} &
\multicolumn{1}{c}{} \\

\noalign{\smallskip}\hline\noalign{\smallskip}
Cen~X-3          &    4.82   &    -(1$-$5)\expo{-12}     &  \citet{1997ApJS..113..367B} \\
SXP\,8.80        &    8.90   &    -5.1\expo{-13}         &  \citet{2010MNRAS.401..252C} \\
GX~1+4           &    110.2  &    -2.1\expo{-13}         &  \citet{1988Natur.333..746M} \\
SXP\,144         &    144.5  &    -5.9\expo{-13}         &  \citet{2010MNRAS.401..252C} \\
SXP\,1062        &    1062   &    -2.6\expo{-12}         &  this work \\
SXP\,1323        &    1325   &    -5.0\expo{-13}         &  \citet{2010MNRAS.401..252C} \\
\noalign{\smallskip}\hline
\end{tabular}
\end{center}
\label{tab-fdot}
\end{table}


We confirm the existence of a new SNR around the BeXRB \sxp in the SMC, independently suggested by \citet{Henault}. 
The SNR is detected in optical, radio, and X-ray wavelengths and shows a shell-like structure with the 
BeXRB close to its projected centre. We estimate
an age of around 10$-$25 ky for the remnant. Such a time scale is too short to spin down the neutron 
star in a BeXRB to $\sim$1000\,s if it was born with a canonical spin period of a few 10 ms. 
\sxp is remarkable for showing a strong average spin-down rate \fdot\ of -2.6\expo{-12} Hz s$^{-1}$ 
during the XMM-Newton observations distributed over 18 days, while it was detected during high X-ray activity 
at 6.3\ergs{35}. Assuming the extreme case that such a high reverse accretion torque was exerted on the 
neutron star for its whole life, a lower limit of 0.5~s for the birth spin period is inferred.
If the neutron star in \sxp is indeed the compact remnant from the supernova explosion that created the SNR,
we conclude that the neutron star in \sxp was born with an even longer spin period for more realistic, lower 
long-term average accretion torques.

\begin{acknowledgements}
The XMM-Newton project is supported by the Bundesministerium f\"ur Wirtschaft und Technologie/Deutsches Zentrum f\"ur 
Luft- und Raumfahrt (BMWI/DLR, FKZ 50 OX 0001) and the Max-Planck Society. 
The MCELS data are provided by R.~C. Smith, P.~F. Winkler, and S.~D. Points. 
The MCELS project has been supported in part by NSF grants AST-9540747 and AST-0307613, and through the generous support 
of the Dean B. McLaughlin Fund at the University of Michigan, a bequest from the family of Dr. Dean B. McLaughlin in 
memory of his lasting impact on Astronomy. The National Optical Astronomy Observatory is operated by the Association of 
Universities for Research in Astronomy Inc. (AURA), under a cooperative agreement with the National Science Foundation.                                                                             
We used the {\sc karma} and {\sc miriad} software package developed by the ATNF. The ATCA is part of the Australia 
Telescope, which is funded by the Commonwealth of Australia for operation as a National Facility managed by CSIRO.
R.S. acknowledges support from the BMWI/DLR grant FKZ 50 OR 0907. 
\end{acknowledgements}

\bibliographystyle{aa}
\bibliography{general}

\end{document}